\documentclass[aip,preprint]{revtex4-1}
\usepackage{graphicx}
\usepackage{amsmath,amssymb}
\usepackage{mathtools}
\usepackage{enumerate}
\usepackage[colorlinks=true,allcolors=blue,breaklinks=true]{hyperref}
\newcommand\revision[1]{#1}

\begin{document}

\title{Diffusion in heterogeneous discs and spheres: new closed-form expressions for exit times and homogenization formulae}
\author{Elliot J. Carr}
\email[Corresponding author: ]{elliot.carr@qut.edu.au}
\affiliation{School of Mathematical Sciences, Queensland University of Technology, Brisbane, Australia.}
\author{Jacob M. Ryan}
\affiliation{School of Mathematical Sciences, Queensland University of Technology, Brisbane, Australia.}
\author{Matthew J. Simpson}
\affiliation{School of Mathematical Sciences, Queensland University of Technology, Brisbane, Australia.}


\begin{abstract}
Mathematical models of diffusive transport underpin our understanding of chemical, biochemical and biological transport phenomena.  Analysis of such models often focusses on relatively simple geometries and deals with diffusion through highly idealised homogeneous media.  In contrast, practical applications of diffusive transport theory inevitably involve dealing with more complicated geometries as well as dealing with heterogeneous media. One of the most fundamental properties of diffusive transport is the concept of \textit{mean particle lifetime} or \textit{mean exit time}, which are particular applications of the concept of \textit{first passage time}, and provide the mean time required for a diffusing particle to reach an absorbing boundary. Most formal analysis of mean particle lifetime applies to relatively simple geometries, often with homogeneous (spatially-invariant) material properties.  In this work, we present a general framework that provides exact mathematical insight into the mean particle lifetime, and higher moments of particle lifetime, for point particles diffusing in heterogeneous discs and spheres with radial symmetry. Our analysis applies to geometries with an arbitrary number and arrangement of distinct layers, where transport in each layer is characterised by a distinct diffusivity.  We obtain exact closed-form expressions for the mean particle lifetime for a diffusing particle released at an arbitrary location and we generalise these results to give exact, closed-form expressions for any higher-order moment of particle lifetime for a range of different boundary conditions. Finally, using these results we construct new homogenization formulae that provide an accurate simplified description of diffusion through heterogeneous discs and spheres.\end{abstract}


\maketitle 

\section{Introduction}
Mathematical models describing diffusive transport of mass and energy are essential for our understanding of many processes in physics~\cite{Redner2001,Krapivsky2010,Hughes1995}, engineering~\cite{Bear1972,Crank1975,Bird2002} and biology~\cite{Murray2002,Codling2008}.  Analysis of mathematical models of diffusive transport  primarily focus on diffusion in relatively simple geometries and homogeneous materials~\cite{Redner2001,Krapivsky2010,Hughes1995,Bear1972,Crank1975,Bird2002}.  In contrast, applications of diffusive transport theory in more complicated geometries and/or with heterogeneous materials are more often explored computationally~\cite{Oran2001,Saxton1994,Lepzelter2012,Ellery2014,Ellery2016,Simpson2018}.  While computational approaches for understanding and interpreting mathematical models of diffusive transport are necessary in certain circumstances, analytical insight is always attractive where possible because it provides simple, easy-to-evaluate, closed-form mathematical expressions that explicitly highlight key relationships~\cite{Simpson2015}.  Such general insight is not always possible when relying on computational methods alone.

A fundamental property of diffusive transport is the concept of \textit{particle lifetime}, which is a particular application of the first passage time~\cite{Redner2001,Krapivsky2010,Hughes1995}.  Developing analytical and computational tools to characterise particle lifetime provides  insight into how varying material properties and geometry affect the time taken for a diffusing particle to reach a certain target~\cite{Lotstedt2015,Meinecke2016a,Meinecke2016b,Meinecke2017,Berezhkovskii2010,Berezhkovskii2011,Carr2018a,Carr2019}. Many results about particle lifetime for diffusive transport have been presented, often in relatively simple homogeneous geometries~\cite{Redner2001,Ellery2012a,Ellery2012b} with certain limited extensions to cases involving more detailed geometries and specific forms of material heterogeneity~\cite{Carr2019,Carr2018b,Kurella2014,Lindsay2015,Vaccario2015}.

In this work, we consider diffusive transport in heterogeneous materials in two and three dimensional domains with radial symmetry.  Such geometries are relevant to a number of important applications in the biophysics literature including the study of transport phenomena in compound droplets~\cite{Landman1983,Landman1985} and the study of nutrient delivery in three dimensional organoid culture~\cite{Ma2013,King2019,Simpson2012,Leedale2019}.   Our modelling framework is very general:  we consider diffusion in discs and spheres, with an arbitrary number and an arbitrary arrangement of distinct layers, where the transport in each layer is characterised by a distinct arbitrary diffusivity.  In this work we show how to obtain exact solutions for the mean particle lifetime for a diffusing particle released at an arbitrary location and we generalise these results to give exact, closed-form expressions for any higher-order moment of particle lifetime for a range of different boundary conditions.   With this information we construct new homogenization formulae~\cite{Davit2013} that allow us to capture key particle lifetime properties in a complex heterogeneous medium with a simpler equivalent homogeneous medium.  These formulae extend many previous formulae that are relevant in one dimension for relatively simple forms of heterogeneity~\cite{Derrida1982,Berezhkovskii2003,Kalnin2013,Kalnin2015,Huysmans2007,Carr2019}.  To test the veracity of the new exact calculations, we implement a stochastic random walk model and show that the exact calculations match appropriately averaged simulation data\cite{SuppMaterial}.  Matlab code to implement the random walk and Maple code to implement the exact calculations are provided on GitHub\cite{Code}.

\begin{figure*}
\includegraphics[width=\textwidth]{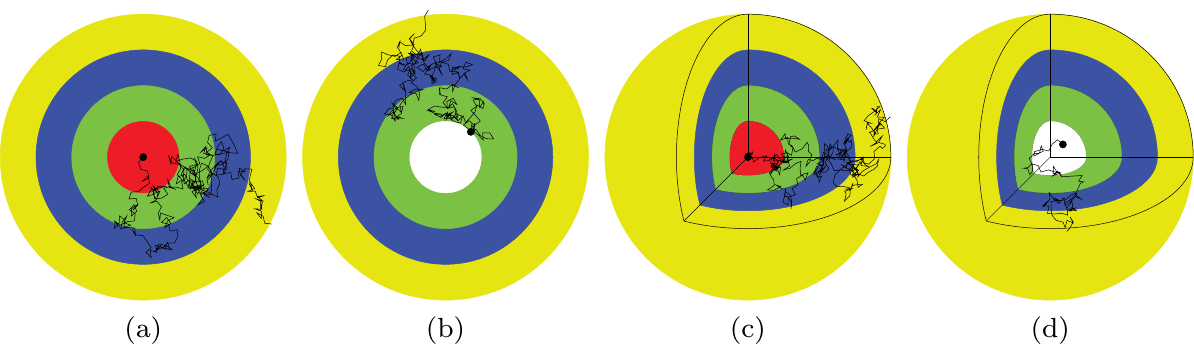}
\caption{Schematic of a random walk for a particle starting at the inner boundary and exiting at the outer boundary for the (a) heterogeneous disc (b) heterogeneous annulus (c) heterogeneous sphere and (d) heterogeneous spherical shell. In each case the inner and outer boundaries are at $r = R_{0}$ and $r = R_{m}$, and the location of the interface between layers $i$ and $i+1$ is $r = R_{i}$ for $i = 1,\hdots, m-1$.  In (a) and (c) $R_{0}=0$, while in (b) and (d) $R_{0}>0$. For the spheres, an octant has been removed to show the layered structure.}
\label{fig:1}
\end{figure*}

\section{Result and Discussion}
\subsection{Discrete model and stochastic simulations}
\label{sec:discrete_model}
Consider a diffusing particle in a line, disc or sphere that is partitioned into $m$ distinct  layers: $\mathcal{L}_{i} = (R_{i-1},R_{i})$ for $i = 1,\hdots,m$ where $R_{0} < R_{1} < \cdots < R_{m}$ (Figure~\ref{fig:1}).  Each layer may take on a distinct diffusivity, $D_i>0$ for $i = 1,\hdots,m$.  The inner and outer boundaries are located at $r = R_{0}$ and $r = R_{m}$, respectively, and $r = R_{i}$ specifies the location of the interface between layers $i$ and $i+1$ ($i = 1,\hdots,m-1$). Note that choosing $R_{0} = 0$ means that we are considering an entire disc or sphere while choosing $R_{0} > 0$ produces an annulus in two dimensions or a spherical shell in three dimensions. If $R_{0} > 0$, either the inner or outer boundary is designated as the absorbing boundary with the other boundary assumed to be reflecting. Otherwise, if $R_{0} = 0$, the outer boundary is designated as the absorbing boundary. We refer to the case of an absorbing outer boundary as the \textit{outward} configuration and the case of an absorbing inner boundary as the \textit{inward} configuration.

We now consider a random walk on the line, disc or sphere.  Here, a particle undergoes a random walk with constant steps of distance $\delta > 0$ and constant time steps of duration $\tau > 0$. When the geometry is heterogeneous: the probability of the particle moving to a new position at each time step varies across the layers with probability $P_{i}$ and diffusivity $D_{i} = P_{i}\delta^{2}/(2d\tau)$ associated with layer $i$, where $d=1,2,3$ is the dimension. For all geometries, the random walk continues until the particle hits the absorbing boundary, at which point the number of steps is recorded. In the following sections, we consider the cases of the disc and sphere only with our implementation of the random walk on a heterogeneous line presented in our previous work\cite{Carr2019}.

\subsubsection{Heterogeneous disc}
\label{sec:heterogeneous_disc}
Let $\mathbf{x}(t) = (x(t),y(t))$ be the position of the particle at time $t$ and $\mathcal{C}(\mathbf{x}(t); \delta)$ be the circle of radius $\delta$ centered at $\mathbf{x}(t)$. If $\mathcal{C}(\mathbf{x}(t); \delta)$ is located entirely within a single layer, say layer $i$ (i.e., $\mathcal{C}(\mathbf{x}(t);\delta)\subset\mathcal{L}_{i}$), then the following outcomes are possible during the next time step: (i) the particle moves to position $\mathbf{x}(t+\tau) = (x(t)+\delta\cos(\theta),y(t)+\delta\sin(\theta))$ with probability $P_{i}$, where $\theta\sim\mathcal{U}[0,2\pi]$; (ii) the particle remains at its current position, $\mathbf{x}(t)$, with probability $1-P_{i}$. If $\mathcal{C}(\mathbf{x}(t);\delta)$ intersects an interface, say $r = R_{i}$, then the following outcomes are possible during the next time step: (i) the particle moves to one of $n$ positions:  $\mathbf{x}(t+\tau) = (x(t)+\delta\cos(\theta_{k}),y(t)+\delta\sin(\theta_{k}))$ where $\theta_{k} = 2\pi (k-1)/n$ with probability $\mathcal{P}_{k}/n$; (ii) the particle remains at its current position, $\mathbf{x}(t)$, with probability $1-\sum_{k=1}^{n}\mathcal{P}_{k}/n$. Here, $\mathcal{P}_{k}$ is the probability associated with the layer in which the position $(x(t)+(\delta/2)\cos(\theta_{k}),y(t)+(\delta/2)\sin(\theta_{k}))$ is located. If $\mathcal{C}(\mathbf{x}(t);\delta)$ intersects the reflecting boundary, the following outcomes are possible during the next time step: (i) the particle attempts to move to position $\mathbf{x}(t+\tau) = (x(t)+\delta\cos(\theta),y(t)+\delta\sin(\theta))$ where $\theta\sim\mathcal{U}[0,2\pi]$ with probability $\mathcal{P}_{b}$; (ii) the particle remains at its current position, $\mathbf{x}(t)$, with probability $1-\mathcal{P}_{b}$. Here, $\mathcal{P}_{b} = P_{1}$ for the outward configuration and $\mathcal{P}_{b} = P_{m}$ for the inward configuration. If the potential step in (i) would require the particle to pass through the reflecting boundary then the step is aborted.

\subsubsection{Heterogeneous sphere}
\label{sec:heterogeneous_sphere}
Let $\mathbf{x}(t) = (x(t),y(t),z(t))$ be the position of the particle at time $t$ and $\mathcal{S}(\mathbf{x}(t); \delta)$ be the sphere of radius $\delta$ centered at $\mathbf{x}(t)$. If $\mathcal{S}(\mathbf{x}(t); \delta)$ is located entirely within a single layer, say layer $i$ (i.e., $\mathcal{S}(\mathbf{x}(t);\delta)\subset\mathcal{L}_{i}$), then the following outcomes are possible during the next time step: (i) the particle moves to position $\mathbf{x}(t+\tau) = (x(t)+\delta\sin(\phi)\cos(\theta),y(t)+\delta\sin(\phi)\sin(\theta),z(t)+\delta\cos(\phi))$ with probability $P_{i}$, where $\phi = \cos^{-1}(1-2u)$, $u\sim\mathcal{U}[0,1]$ and $\theta\sim\mathcal{U}[0,2\pi]$; (ii) the particle remains at its current position, $\mathbf{x}(t)$, with probability $1-P_{i}$. The formula for $\phi$ avoids the clustering of random points around the poles ($\phi = 0$ and $\phi = \pi$) when $\phi$ is naively sampled from $\mathcal{U}[0,\pi]$ \cite{Weisstein}. If $\mathcal{S}(\mathbf{x}(t);\delta)$ intersects an interface, say $r = R_{i}$, then the following outcomes are possible during the next time step: (i) the particle moves to one of $n = n_{1}n_{2}$ positions: $\mathbf{x}(t+\tau) = (x(t)+\delta\sin(\phi_{j})\cos(\theta_{k}),y(t)+\delta\sin(\phi_{j})\sin(\theta_{k}),y(t)+\delta\cos(\phi_{j}))$ ($j = 1,\hdots,n_{1}$, $k = 1,\hdots,n_{2}$) with probability $\mathcal{P}_{j,k}/n$, where $\phi_{j} = \cos^{-1}(1-2(j-1)/n_{1})$ and $\theta_{k} = 2\pi (k-1)/n_{2}$; (ii) the particle remains at its current position, $\mathbf{x}(t)$, with probability $1-\sum_{j=1}^{n_{1}}\sum_{k=1}^{n_{2}}\mathcal{P}_{j,k}/n$. Here, $\mathcal{P}_{j,k}$ is the probability associated with the layer in which the position $(x(t)+(\delta/2)\sin(\phi_{j})\cos(\theta_{k}),y(t)+(\delta/2)\sin(\phi_{j})\sin(\theta_{k}),y(t)+\delta\cos(\phi_{j}))$ is located. If $\mathcal{S}(\mathbf{x}(t);\delta)$ intersects the reflecting boundary, the following outcomes are possible during the next time step: (i) the particle attempts to move to position $\mathbf{x}(t+\tau) = (x(t)+\delta\sin(\phi)\cos(\theta),y(t)+\delta\sin(\phi)\sin(\theta),z(t)+\delta\cos(\phi))$ with probability $\mathcal{P}_{b}$, where $\phi = \cos^{-1}(1-2u)$, $u\sim\mathcal{U}[0,1]$ and $\theta\sim\mathcal{U}[0,2\pi]$; (ii) the particle remains at its current position, $\mathbf{x}(t)$, with probability $1-\mathcal{P}_{b}$. Here, $\mathcal{P}_{b} = P_{1}$ for the outward configuration and $\mathcal{P}_{b} = P_{m}$ for the inward configuration. If the potential step in (i) would require the particle to pass through the reflecting boundary then the step is aborted.

MATLAB implementations of the random walk on the heterogeneous disc and sphere are available on GitHub\cite{Code} and simulation results are summarised in the Supplementary Material~\cite{SuppMaterial}.  These algorithms are constructed so that we can consider diffusive transport in two or three dimensions with arbitrary choices of $R_{0} < R_{1} < \cdots < R_{m}$ and $D_i>0$ for $i = 1,\hdots,m$. The code accommodates both the outward ($r = R_{m}$ absorbing) and inward ($r = R_{0}$ absorbing) configurations as well as allowing for the initial location of the particle to be chosen arbitrarily in the interval $[R_{0},R_{m}]$. To provide mathematical insight into these simulations we now consider analysing the moments of exit time.

\subsection{Moments of exit time}
\label{sec:moments}
Due to the symmetries inherent in the heterogeneous disc and sphere, the exit time properties are independent of the angles $\theta$ and $\phi$ and depend only on the radial coordinate $r$. Let $\mathbb{E}(T^{k}; r)$ be the $k$th moment of exit time for a particle with starting position $x(0) = r$ (line), $\mathbf{x}(0) = (x(0),y(0)) = (r,0)$ (disc) and $\mathbf{x}(0) = (x(0),y(0),z(0)) = (r,0,0)$ (sphere). Suppose for either the line, disc or sphere that the random walk is repeated $n$ times with the same starting location $r$ yielding the following recorded exit times: $T_{1},T_{2},\hdots,T_{n}$. Then the $k$th raw moment of exit time is estimated using the stochastic simulations via:
\begin{gather}
\label{eq:stochastic_moment}
\mathbb{E}(T^{k}; r) \approx \frac{1}{n}\sum_{j=1}^{n} T_{j}^{k},
\end{gather}
with equality obtained in the limit as $n\rightarrow\infty$. \revision{Using our previous arguments\cite{Carr2019}, we confirm that the continuum representation of $\mathbb{E}(T^{k}; r)$ in layer $i$, which we denote by $M_{k}^{(i)}(r)$, satisfies the following system of differential equations\cite{Redner2001,Karlin1981}}:
\begin{gather}
\label{eq:moment_ode}
\dfrac{D_{i}}{r^{d-1}}\dfrac{\textrm{d}}{\textrm{d}r}\left(r^{d-1}\dfrac{\textrm{d}M_{k}^{(i)}}{\textrm{d} r}\right) = -kM_{k-1}^{(i)},
\end{gather}
for $i = 1,\hdots,m$ where $r\in(R_{i-1},R_{i})$ and $d = 1,2,3$ is the dimension. The appropriate internal boundary conditions at the interfaces are
\begin{gather}
\label{eq:moment_ode_interface1}
M_{k}^{(i-1)}(R_{i}) = M_{k}^{(i)}(R_{i}),\\
\label{eq:moment_ode_interface2}
D_{i-1}\frac{\text{d}M_{k}^{(i-1)}}{\text{d}r}(R_{i}) = D_{i}\frac{\text{d}M_{k}^{(i)}}{\text{d}r}(R_{i}),
\end{gather}
for $i = 1,\hdots,m-1$. These equations must also be supplemented with boundary conditions at $r = R_{0}$ and $r = R_{m}$. The appropriate conditions are $\textrm{d}M_{k}^{(1)}(R_{0})/\textrm{d}r = 0$ and $M_{k}^{(m)}(R_{m}) = 0$ for the outward configuration and $M_{k}^{(1)}(R_{0}) = 0$ and $\textrm{d}M_{k}^{(m)}(R_{m})/\textrm{d}r = 0$ for the inward configuration. For the disc and sphere, if $R_{0}=0$ then the inner boundary vanishes and a symmetry condition is imposed, $\textrm{d}M_{k}^{(1)}(0)/\textrm{d}r = 0$. The boundary value problem for $M_{k}^{(i)}$ for $i = 1,\hdots,m$ is solved sequentially for $k = 1,2,\hdots$ given $M_{0}^{(i)}(r) = 1$ for all $r\in (R_{i-1},R_{i})$ and $i = 1,\hdots,m$, which is evident from   (\ref{eq:stochastic_moment}) when $k = 0$.

The attraction of working with the moments of exit time is that the system of boundary value problems can be solved exactly to obtain closed-form analytical expressions for the moments in each layer in terms of the diffusivities $D_{1},\hdots,D_{m}$ and radii $R_{0},\hdots,R_{m}$. This can be achieved using standard symbolic software. A Maple worksheet that is capable of solving these boundary value problems is available on GitHub\cite{Code} and a comparison between appropriately averaged data from the stochastic simulation algorithm and the solution of the relevant boundary value problems that confirm the accuracy of the expressions for the moments of exit time are detailed in the Supplementary Material~\cite{SuppMaterial}.

\begin{table*}[t]
	\centering
	\caption{Moments of exit time formulae for homogeneous ($m=1$) and heterogeneous ($m > 1$) discs and spheres with $R_0=0$ under the outward configuration. For the homogeneous case, we drop the layer index on the moments appearing in the superscript.}
	\label{Table1}
	\begin{tabular}{c | c}
		\hline Homogeneous disc & Homogeneous sphere \\ \hline \hline \\[-0.3cm]
		$\displaystyle M_1(r) = \frac{R_{1}^2 - r^2}{4D_{1}} $  & $\displaystyle M_1(r) = \frac{R_{1}^2 - r^2}{6D_{1}} $\\[0.2cm]
		$\displaystyle M_2(r) = \frac{R_{1}^2 - r^2}{32D_{1}^2}\left( 3R_{1}^2 - r^2 \right) $ & $\displaystyle M_2(r) = \frac{R_{1}^2 - r^2}{180D_{1}^2}\left( 7R_{1}^2 - 3r^2 \right) $\\[0.2cm]
		$\displaystyle M_3(r) = \frac{R_{1}^2 - r^2}{384D_{1}^3}\left( 19R_{1}^4 - 8R_{1}^2r^2 + r^4 \right) $ & $\displaystyle M_3(r) = \frac{R_{1}^2 - r^2}{2520D_{1}^3}\left( 31R_{1}^4 - 18R_{1}^2r^2 + 3r^4 \right) $\\ \\[-0.3cm]
		\hline Heterogeneous disc (2 layers) & Heterogeneous sphere (2 layers) \\ \hline \hline  \\[-0.3cm]
		$\displaystyle M_1^{(1)}(r) = \frac{R_1^2 - r^2}{4D_1} + \frac{R_2^2 - R_{1}^2}{4D_2} $ & $\displaystyle M_1^{(1)}(r) = \frac{R_1^2 - r^2}{6D_1} + \frac{R_2^2 - R_{1}^2}{6D_2} $\\[0.2cm]
		$\displaystyle M_1^{(2)}(r) = \frac{R_2^2 - r^2}{4D_2} $ & $\displaystyle M_1^{(2)}(r) = \frac{R_2^2 - r^2}{6D_2} $\\ \\[-0.3cm]
				\hline Heterogeneous disc ($m$ layers) & Heterogeneous sphere ($m$ layers) \\ \hline \hline  \\[-0.3cm]
		$\displaystyle M_1^{(i)}(r) = \frac{R_i^2 - r^2}{4D_i} + \sum_{j=i+1}^m \frac{R_j^2 - R_{j-1}^2}{4D_j}$, & $\displaystyle M_1^{(i)}(r) = \frac{R_i^2 - r^2}{6D_i} + \sum_{j=i+1}^m \frac{R_j^2 - R_{j-1}^2}{6D_j},$\\ $i = 1,\hdots,m.$ & $i = 1,\hdots,m.$\\[-0.3cm] \\ \hline
	\end{tabular}
\end{table*}

\begin{table*}[t]
	\centering
	\renewcommand*{\arraystretch}{1.0}
	\caption{Moments of exit time formulae for a homogeneous ($m = 1$) or heterogeneous ($m > 1$) annulus ($R_{0}>0$) under the inward or outward configuration. For the homogeneous case, we drop the layer index on the moments appearing in the superscript.}
	\label{Table2}
	\begin{tabular}{c}
		\hline Homogeneous disc (outward configuration) \\ \hline \hline\\[-0.3cm]
		$\displaystyle M_1(r) = \frac{R_{1}^2 - r^2}{4D_{1}} + \frac{R_0^2}{2D_{1}} \ln{\left(  \frac{r}{R_{1}} \right)}$\\
		$\displaystyle M_2(r) = \frac{R_0^2}{32D_{1}^2} \left[8\left( r^2 + \frac{3R_0^2}{2} - R_{1}^2\right)\ln\left(\frac{R_{1}}{r} \right) + 16R_0^2\left( \ln(R_{1})^2 + \ln(R_0) \ln(r) - \ln(R_0 r) \ln(R_{1}) \right) \right] $\\
		$\displaystyle + \frac{R_{1}^2 - r^2}{32D_{1}^2}\left( 3R_{1}^2 - r^2 - 8R_0^2 \right) $ \\ \\[-0.3cm]
		\hline Homogeneous disc (inward configuration) \\ \hline \hline\\[-0.3cm]
		$\displaystyle M_1(r) = \frac{R_0^2 - r^2}{4D_{1}} + \frac{R_{1}^2}{2D_{1}}\ln\left( \frac{r}{R_0} \right)$ \\
		$\displaystyle M_2(r) = \frac{R_{1}^2}{32D_{1}^2} \left[8\left( r^2 + \frac{3R_{1}^2}{2} - R_0^2\right)\ln\left(\frac{R_0}{r} \right) + 16R_{1}^2\left(\ln(R_0)^2 + \ln(R_1) \ln(r) - \ln(R_1 r) \ln(R_0) \right) \right]$\\
		$\displaystyle + \frac{R_0^2 - r^2}{32D_{1}^2}\left( 3R_{0}^2 - r^2 - 8R_1^2 \right) $ \\ \\[-0.3cm]
		\hline Heterogeneous disc ($m$ layers, outward configuration) \\ \hline \hline \\[-0.3cm]
		$\displaystyle M_1^{(i)}(r) = \frac{R_i^2 - r^2}{4D_i} + \frac{R_0^2}{2D_i}\ln \left( \frac{r}{R_i} \right) + \sum_{j=i+1}^m \frac{R_j^2 - R_{j-1}^2}{4D_j} + \frac{R_0^2}{2D_j}\ln \left(\frac{ R_{j-1} }{R_j} \right),$\\ $i = 1,\hdots,m.$ \\ \\[-0.3cm]
		\hline Heterogeneous disc ($m$ layers, inward configuration) \\ \hline \hline\\[-0.3cm]
		$\displaystyle M_1^{(i)}(r) = \frac{R_{i-1}^2 - r^2}{4D_i} + \frac{R_m^2}{2D_i} \ln \left( \frac{r}{ R_{i-1} } \right) + \sum_{j=1}^{i-1} \frac{R_{j-1}^2 - R_{j}^2}{4D_j} + \frac{R_m^2}{2D_j} \ln \left(\frac{ R_{j} }{ R_{j-1} } \right),$\\ $i = 1,\hdots,m.$ \\[-0.3cm] \\ \hline
	\end{tabular}
\end{table*}

\begin{table*}[t]
	\centering
	\caption{Moments of exit time formulae for a homogeneous ($m = 1$) or heterogeneous ($m > 1$) spherical shell ($R_{0}>0$) under the inward or outward configuration. For the homogeneous case, we drop the layer index on the moments appearing in the superscript.}
	\label{Table3}
	\begin{tabular}{c}
		\hline Homogeneous sphere (outward configuration) \\ \hline \hline \\[-0.3cm]
		$\displaystyle M_1(r) = \frac{R_{1}^2 - r^2}{6D_{1}} + \frac{R_0^3}{3D_{1}}\left( \frac{1}{R_{1}} - \frac{1}{r} \right) $ \\
		$\displaystyle M_2(r) = \frac{R_{1}^2 - r^2}{180D_{1}^2}\left( 7R_{1}^2 - 3r^2 \right) + \frac{R_0^3}{D_{1}^2}\left[ \frac{1}{9}\left( 3r - R_{1} - \frac{R_{1}^2}{r} - \frac{r^2}{R_{1}} \right) + \frac{2R_0^2}{5}\left( \frac{1}{r} - \frac{1}{R_{1}} \right) + \frac{2R_0^3}{9R_{1}}\left( \frac{1}{R_{1}} - \frac{1}{r} \right)  \right]  $ \\ \\[-0.3cm]
		\hline Homogeneous sphere (inward configuration) \\ \hline \hline \\[-0.3cm]
		$\displaystyle M_1(r) = \frac{R_0^2 - r^2}{6D_{1}} + \frac{R_{1}^3}{3D_{1}}\left( \frac{1}{R_0} - \frac{1}{r} \right)$ \\
		$\displaystyle M_2(r) = \frac{R_0^2 - r^2}{180D_{1}^2}\left( 7R_0^2 - 3r^2 \right) + \frac{R_{1}^3}{D_{1}^2}\left[ \frac{1}{9}\left( 3r - R_0 - \frac{R_0^2}{r} - \frac{r^2}{R_0} \right) + \frac{2R_{1}^2}{5}\left( \frac{1}{r} - \frac{1}{R_0} \right) + \frac{2R_{1}^3}{9R_0}\left( \frac{1}{R_0} - \frac{1}{r} \right)  \right]  $ \\ \\[-0.3cm]
		\hline Heterogeneous sphere ($m$ layers, outward configuration) \\ \hline \hline \\[-0.3cm]
		$\displaystyle M_1^{(i)}(r) = \frac{R_i^2 - r^2}{6D_i} + \frac{R_0^3}{3D_i}\left[ \frac{1}{R_i} - \frac{1}{r} \right] + \sum_{j=i+1}^m \frac{R_j^2 - R_{j-1}^2}{6D_j} + \frac{R_0^3}{3D_j}\left[ \frac{1}{R_j} - \frac{1}{R_{j-1}} \right],$\\ $i = 1,\hdots,m.$\\ \\[-0.3cm]
		\hline Heterogeneous sphere ($m$ layers, inward configuration) \\ \hline \hline \\[-0.3cm]
		$\displaystyle M_1^{(i)}(r) = \frac{R_{i-1}^2 - r^2}{6D_i} + \frac{R_m^3}{3D_i}\left[ \frac{1}{R_{i-1}} - \frac{1}{r} \right] + \sum_{j=1}^{i-1} \frac{R_{j-1}^2 - R_{j}^2}{6D_j} + \frac{R_m^3}{3D_j}\left[ \frac{1}{R_{j-1}} - \frac{1}{R_{j}} \right],$\\ $i = 1,\hdots,m.$\\[-0.3cm] \\\hline
	\end{tabular}
\end{table*}

A summary of closed-form expressions is provided in Table~\ref{Table1} for the outward configuration and the case where $R_{0}=0$.  The first three rows in Table~\ref{Table1} show the first three moments for an homogeneous medium ($m=1$) and we see that the algebraic expressions become increasingly complicated as we consider higher moments.  The fourth row in Table~\ref{Table1} shows simple expressions for the first moment of exit time in a two layer problem.  Experimentation with Maple enables us to propose a more general expression for the first moment of exit time in a more general scenario with $m$ layers, as shown in the fifth row of Table~\ref{Table1}.  While it is possible to use the Maple worksheet provided to find values of an arbitrary moment of exit time in a problem with an arbitrary number of layers, it is much more difficult to give closed-form expressions for such higher moments.

All expressions in Table~\ref{Table1} are for discs and spheres with $R_{0}=0$ under the outward configuration.  Additional results in Tables~\ref{Table2}--\ref{Table3} compare moments of exit time formulae for an annulus and spherical shell ($R_{0}>0$) under both the outward and inward configurations. Here we see a key difference between results for one dimension with results in higher dimensions. For example, consider the simplest possible case of a random walk on a finite line ($d=1$) in a homogeneous environment.  If a diffusing particle is released in the centre of the domain, the expected time to reach either boundary is equal~\cite{Hughes1995,Redner2001}.  In contrast if we have a random walk in a radially symmetric homogeneous disc or sphere and a particle is released at the centre of the domain, the diffusing particle takes different amounts of time to reach the two boundaries owing to the radial geometry.  Such differences can be even more nuanced when the domain is heterogeneous and such considerations can have practical implications.   For example, consider the case where we treat a biological cell as a two-layer compound sphere ($m=2$), with the outer layer ($R_1 < r <R_2$) representing the cytoplasm and the inner layer ($0 < r <R_1$) representing the nucleus.  In this case it is of interest to estimate the expected time taken for a molecule released at the outer surface ($r=R_2$) to be absorbed at the nuclear membrane ($r=R_1$).  Alternatively, it is also of interest to estimate the expected time taken for a protein synthesized in the nucleus to diffuse from the nuclear membrane ($r=R_1$) to the exterior cell membrane ($r=R_2$). These effects of directionality are explicitly described in the formulae in Tables~\ref{Table2}--\ref{Table3}. 

Vaccario et al.\cite{Vaccario2015} present closed-form expressions for the first moment of exit time for lines, discs and spheres with $R_{0} > 0$ derived using a stochastic differential equation description of the random walk process. Results are given for two layers under the inward configuration and depend on the convention of the stochastic integral (It\^o, Stratonovich or isothermal). Interestingly, our closed-form expressions for the first moment of exit time in Table \ref{Table3} match those reported by Vaccario et al. for the isothermal convention. \revision{As reported in the Supplementary Material\cite{SuppMaterial}, the solution of the continuum description (\ref{eq:moment_ode})--(\ref{eq:moment_ode_interface2}) matches with appropriately averaged data from our discrete stochastic model, which includes an assumption specifying how the random walk interacts with the interfaces at $r = R_{i}$ for $i = 1,\hdots,m-1$ (see Sections \ref{sec:heterogeneous_disc}--\ref{sec:heterogeneous_sphere}). We acknowledge that a different assumption at the interface would likely lead to a different continuum description. Nonetheless we focus here on this particular implementation since our assumptions about the jump probabilities at the interface are minimal and provide a straightforward means of incorporating spatial variations in diffusivity.} 

\begin{figure*}[t]
	\centering
	\includegraphics[width=1.0\textwidth]{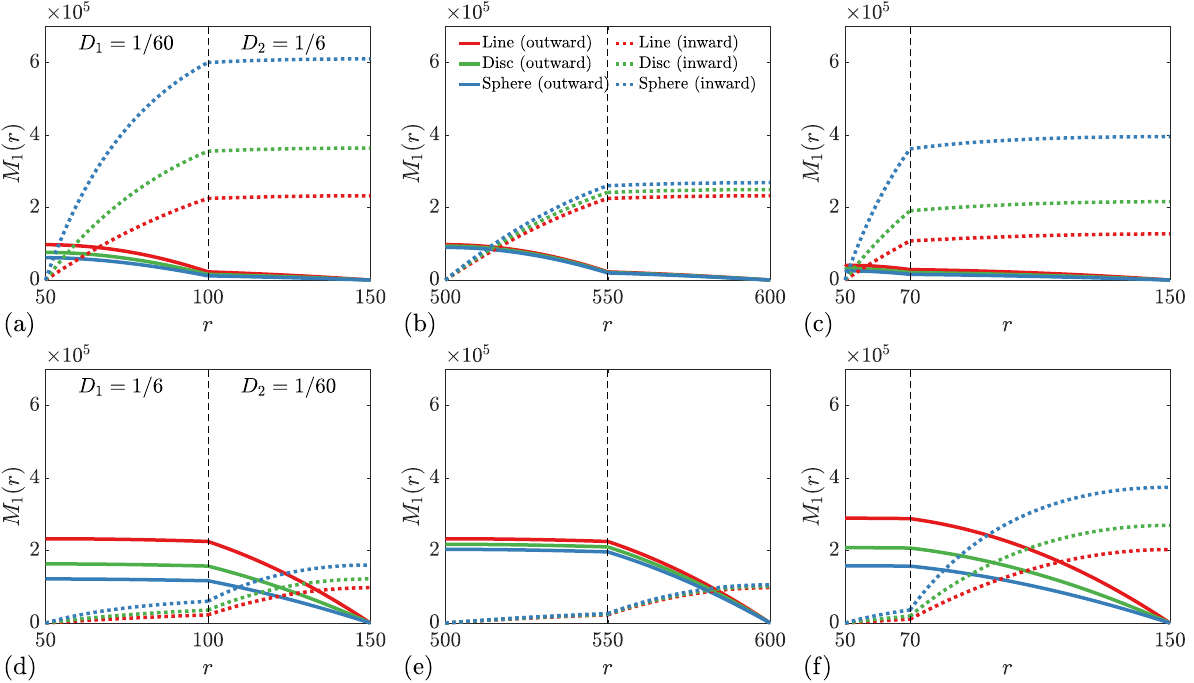}
	\caption{First moment of exit time $M_1(r)$ for a range of cases in heterogeneous discs and spheres with two layers.  Each subfigure compares random walks on lines ($d=1$, red), discs ($d=2$, green) and spheres ($d=3$, blue) for both the outward (solid) and inward (dashed) configurations. For (a)--(c) we have $D_1 = 1/60$ and $D_2 = 1/6$, while for (d)--(f) we have $D_1 = 1/6$ and $D_2 = 1/60$.  Cases in (a), (c), (d) and (f) correspond to $R_{0}=50$ and $R_{2}=150$, while cases in (b) and (e) correspond to $R_{0}=500$ and $R_{2}=600$. The legend in (b) applies to all subfigures.}
	\label{fig:2}
\end{figure*}

Figure~\ref{fig:2} compares the spatial distribution of the first moment of exit times for a range of problems in heterogeneous lines, discs and spheres with two layers ($m=2$).  Profiles in Figure~\ref{fig:2}(a) show the first moment for a problem with $50 < r < 150$ with an interface at the midpoint, $r=100$.  Results for the outward configuration show that the spatial distribution of the first moment is relatively sensitive to the dimension of the problem since we observe distinct results for $d=1,2,3$.  Similarly, results for the inward configuration indicate that the spatial distribution of the first moment is also relatively sensitive to the dimensionality. Results in Figure \ref{fig:2}(a) demonstrate that for the inward configuration, the first moment of exit time is largest for the sphere and smallest for the line while the opposite is true for the outward configuration. For example, the mean time for a particle released at the outer boundary to exit at the inner boundary for the problem in Figure \ref{fig:2}(a) is $2.3\times 10^{5}$, $3.6\times 10^{5}$ and $6.1\times 10^{5}$ for the line, disc and sphere, respectively. On the other hand, the mean time for a particle released at the inner boundary to exit at the outer boundary is $9.8\times 10^{4}$, $7.6\times 10^{4}$ and $6.2\times 10^{4}$ for the line, disc and sphere, respectively. These differences can be explained by the geometry-induced outward bias inherent in the random walk on the disc and the sphere, where a particle is more likely to move away from the origin in the positive $r$ direction due to the circular and spherical geometries. This geometry-induced outward bias is absent on the line, which explains why the mean exit time on a line is smallest for the inward configuration and largest for the outward configuration (see Figure \ref{fig:2}(a)).

Results in Figure~\ref{fig:2}(b) show the spatial distribution of the first moment for a similar problem with $500 < r < 600$ with an interface at the midpoint, $r=550$.  Therefore, the dimensions of the problem in Figure~\ref{fig:2}(b) are very similar to those in Figure~\ref{fig:2}(a) except the effects of the geometry are more pronounced in Figure~\ref{fig:2}(a) since the domain is closer to the origin.  As a result, the profiles for the first moment in Figure~\ref{fig:2}(b) are much less sensitive to position than those in Figure~\ref{fig:2}(a) since the geometric differences between $d=1,2$ and $d=3$ are far less pronounced when $r$ is larger. This is explained by the geometry-induced outward bias present for the disc and sphere, which is proportional to $(d-1)/r$ and therefore more dominant when the domain is closer to the origin. This observation also explains why the spatial distributions of the first moment of exit time for the disc and sphere approach the spatial distributions of the first moment of exit time for the line in Figure \ref{fig:2}(b) since the domain $500<r<600$ is far from the origin. For the outward configuration, the spatial distributions of the first moment of exit time for the disc and sphere increase from Figure \ref{fig:2}(a) to Figure \ref{fig:2}(b) since a particle is less likely to move outwards for the problem in Figure \ref{fig:2}(b) than for the problem in Figure \ref{fig:2}(a). Similarly, for the inward configuration, the first moment of exit time for the disc and sphere decrease since a particle is more likely to move inwards for the problem in Figure \ref{fig:2}(b) than for the problem in Figure \ref{fig:2}(a). The spatial distributions of the first moment of exit time for the line in Figure \ref{fig:2}(b) remain unchanged from Figure \ref{fig:2}(a) as there is no geometry-induced outward bias for the random walk on the line. 

Profiles in Figure~\ref{fig:2}(c) show the first moment for a problem with $50 < r < 150$ with an interface at $r=70$.  Therefore, the only difference between the problems in Figure~\ref{fig:2}(a) and Figure~\ref{fig:2}(c) is the location of the interface and here we see that when the interface is at $r=70$ the dependence on $d$ is much less pronounced for the outward configuration whereas the dependence on $d$ is much more pronounced for the inward configuration. For the inward configuration in Figure \ref{fig:2}(c), a particle released in the second layer is required to pass through a thinner layer of the lower diffusivity material to exit at $r = 50$ compared to Figure \ref{fig:2}(a). This explains the results in Figure \ref{fig:2}(a) and Figure \ref{fig:2}(c) for the inward configuration, where we see that the mean exit time remains largely unchanged for $50<r<70$ but reduces significantly for $70<r<150$.

Results in Figures~\ref{fig:2}(d)--(f) show the spatial distribution of the first moment when the values of $D_{1}$ and $D_{2}$ for the problems in Figures~\ref{fig:2}(a)--(c) are interchanged. Comparing the results for the inward configuration in Figures \ref{fig:2}(a)--(b) with the results for the outward configuration in Figures \ref{fig:2}(d)--(e), we see that the profiles for the line are symmetrical to each other about the interface. As there is no geometry-induced outward bias on the line and the interface is located at the midpoint of the domain, the outward configuration in Figures \ref{fig:2}(a)--(b) is equivalent to the inward configuration in Figures \ref{fig:2}(d)--(e) leading to the observed symmetry. Another observation evident from Figure \ref{fig:2}(d)--(f) is that the first moment of exit time is approximately uniform in the first layer for the outward configuration. This is because $D_{1}$ is an order of magnitude greater than $D_{2}$. Thus, the exit time for a particle released in the first layer is dominated by the time taken for the particle to pass through the second layer. A similar argument explains the approximately uniform mean exit time distribution in the second layer for the inward configuration in Figure \ref{fig:2}(a)--(c).

The key trends in the distribution of the first moment profiles in Figure~\ref{fig:2}(a)--(f) are echoed in the trends in the second and higher moment profiles (not shown).  Now that we have constructed, validated~\cite{SuppMaterial}, and explored exact solutions for arbitrary moments of exit time for diffusion in a wide range of heterogeneous discs and spheres we can use these results to derive new homogenization formulae.

\subsection{Homogenization results}
We now consider the problem of approximating a stochastic random walk on a heterogeneous disc or sphere with a stochastic random walk on an equivalent or \textit{effective} homogeneous disc or sphere. Here we assume the homogenized stochastic model takes the form of an unbiased random walk with diffusivity $D_{\mathrm{eff}}$ and probability $P_{\mathrm{eff}} = 2d\tau D_{\mathrm{eff}}/\delta^{2}$.  To keep the derivation succinct we consider the case where the particles are released at the reflecting boundary ($r = R_{0}$ for the outward configuration and $r = R_{m}$ for the inward configuration).  We choose $D_{\mathrm{eff}}$ to constrain the first moment of exit time for the homogenized random walk to be equal to the first moment of exit time for the heterogeneous random walk at the starting position of the particle, that is $M_1^{\mathrm{eff}}(R_{0}) = M_1^{(1)}(R_{0})$ for the outward configuration and $M_1^{\mathrm{eff}}(R_{m}) = M_1^{(m)}(R_{m})$ for the inward configuration. Combining these constraints with the closed-form expressions for the moments of exit time in Tables~\ref{Table1}--\ref{Table3} (note: $M_1^{\mathrm{eff}}(r)$ denotes $M_{1}(r)$ with $D_{1} = D_{\mathrm{eff}}$ and $R_{1} = R_{m}$) and rearranging for $D_{\mathrm{eff}}$ yields simple closed-form formulae for the effective diffusivity listed in Table~\ref{Table4}. Identifying the pattern in these homogenization formulae and the corresponding formulae for the heterogeneous line\cite{Carr2019}, we identify the general homogenization formula given in Table~\ref{Table4}, which is valid for all three dimensions ($d = 1,2,3$) and both the inward and outward configurations for the case when a particle is released at the reflecting boundary.

\begin{table*}[t]
	\centering
	\caption{Effective diffusivity formulae for a heterogeneous ($m > 1$) annulus or spherical shell ($R_{0}>0$) under the inward or outward configuration for a particle released at the reflecting boundary ($r = R_{0}$ for the outward configuration and $r = R_{m}$ for the inward configuration).}
	\label{Table4}
	\begin{tabular}{c}
		\hline Heterogeneous disc ($m$ layers, outward configuration) \\ \hline \hline \\[-0.3cm]
		$\displaystyle D_{\mathrm{eff}} =  \left. \Biggl[\frac{R_{m}^2 - R_{0}^2}{4} + \frac{R_0^2}{2} \ln{\left(  \frac{R_{0}}{R_{m}} \right)}\Biggr] \middle/ \Biggl[\sum_{j=1}^m \frac{R_j^2 - R_{j-1}^2}{4D_j} + \frac{R_0^2}{2D_j}\ln \left(\frac{ R_{j-1} }{R_j} \right)\Biggr] \right.  $ \\ \\[-0.3cm]
		\hline Heterogeneous disc ($m$ layers, inward configuration) \\ \hline \hline \\[-0.3cm]
				$\displaystyle D_{\mathrm{eff}} = \left.  \Biggl[\frac{R_m^2 - R_{0}^2}{4} + \frac{R_{m}^2}{2}\ln\left( \frac{R_0}{R_{m}} \right)\Biggr] \middle / \Biggl[\sum_{j=1}^{m} \frac{R_{j}^2 - R_{j-1}^2}{4D_j} + \frac{R_m^2}{2D_j} \ln \left(\frac{ R_{j-1} }{ R_{j} } \right)\Biggr] \right. $ \\ \\[-0.3cm]
		\hline Heterogeneous sphere ($m$ layers, outward configuration) \\ \hline \hline \\[-0.3cm]
		$\displaystyle D_{\mathrm{eff}} = \left.   \Biggl[\frac{R_{m}^2 - R_{0}^2}{6} + \frac{R_0^3}{3}\left( \frac{1}{R_{m}} - \frac{1}{R_{0}} \right)\Biggr] \middle/ \Biggl[\sum_{j=1}^m \frac{R_j^2 - R_{j-1}^2}{6D_j} + \frac{R_0^3}{3D_j}\left( \frac{1}{R_j} - \frac{1}{R_{j-1}} \right)\Biggr] \right.$ \\ \\[-0.3cm]
		\hline Heterogeneous sphere ($m$ layers, inward configuration) \\ \hline \hline \\[-0.3cm]
$\displaystyle D_{\mathrm{eff}} = \left.  \Biggl[\frac{R_m^2 - R_{0}^2}{6} + \frac{R_{m}^3}{3}\left( \frac{1}{R_m} - \frac{1}{R_{0}} \right)\Biggr]\middle/ \Biggl[\sum_{j=1}^{m} \frac{R_{j}^2 - R_{j-1}^2}{6D_j} + \frac{R_m^3}{3D_j}\left( \frac{1}{R_{j}} - \frac{1}{R_{j-1}} \right)\Biggr] \right.$ \\ \\[-0.3cm] \hline
\hline Heterogeneous line ($d=1$), disc ($d = 2$) or sphere ($d=3$) ($m$ layers) \\ \hline \hline \\[-0.3cm]
		$\displaystyle D_{\mathrm{eff}} = \left.   \Biggl[\frac{R_{m}^2 - R_{0}^2}{2d} - \frac{\widetilde{R}^d}{d}\int_{R_{0}}^{R_{m}}\frac{1}{r^{d-1}}\,\text{d}r\Biggr] \middle/ \Biggl[\sum_{j=1}^m \frac{R_j^2 - R_{j-1}^2}{2dD_j} - \frac{\widetilde{R}^d}{dD_j}\int_{R_{j-1}}^{R_{j}}\frac{1}{r^{d-1}}\,\text{d}r\Biggr] \right.$ \\ \text{where $\widetilde{R} = R_{0}$ (outward) or $\widetilde{R} = R_{m}$ (inward).} \\ \\[-0.3cm] \hline
	\end{tabular}
\end{table*}

\begin{figure*}[t]
	\centering
	\includegraphics[width=1.0\textwidth]{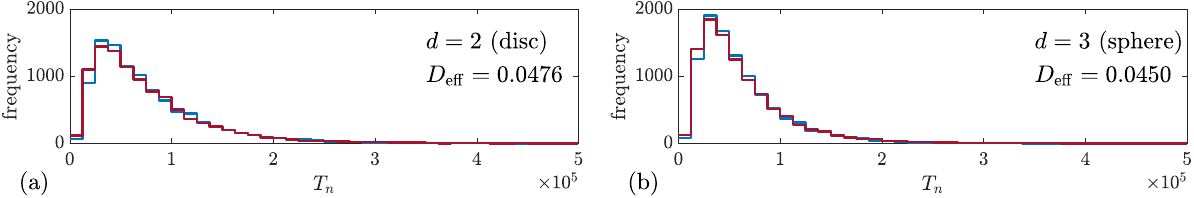}
	\caption{Histograms of 10000 realisations of a two-layer heterogeneous random walk (thick blue) and an effective homogenous random walk (thin red). The two-layer random walk is described by $D_1 = 1/60$, $D_2 = 1/6$, $R_0 = 50$, $R_1 = 100$ and $R_2=150$. The effective homogenous random walk is described using the same geometry but with $D_{1} = D_{2} =  D_{\mathrm{eff}}$ from Table~\ref{Table4}. Both random walks assume an outward configuration and $\delta = \tau = 1$. Results in (a) correspond to $n=24$ and (b) correspond to $n_{1}=n_{2}=12$ (see Sections \ref{sec:heterogeneous_disc}--\ref{sec:heterogeneous_sphere}).}
	\label{fig:3}
\end{figure*}

We now provide a visual interpretation of the homogenization approximation for heterogeneous discs and spheres in Figure~\ref{fig:3}.  The results in this figure are generated by considering a random walk with $\delta = \tau = 1$ in a heterogeneous disc (Figure~\ref{fig:3}(a)) and a heterogeneous sphere (Figure~\ref{fig:3}(b)) consisting of two layers ($m=2$) with $R_{0}=50$, $R_{2}=150$ and the interface at the centre, $R_{1}=100$.  We consider an order-of-magnitude difference in the diffusivities in the two media so that we demonstrate the performance of the new homogenization formulae for relatively strong heterogeneity~\cite{Carr2019}.  For both the disc and sphere, we consider 10,000 identically-prepared outward-configuration simulations with a particle placed at $r=R_{0}$ and simulations performed until the particle reaches the absorbing boundary at $r=R_{2}$.  In each simulation we record the exit time, $T_{n}$ for $n=1,\hdots, 10,000$ and we construct a histogram of the particle lifetime as given in Figure~\ref{fig:3}.

To generate the effective homogenised random walk results in Figure~\ref{fig:3} we simulate using the same geometry and boundary conditions as in the heterogeneous case but with $D_{1} = D_{2} = D_{\mathrm{eff}}$, where $D_{\mathrm{eff}}$ is computed from the outward configuration formulae in Table~\ref{Table4}. Performing 10,000 identically prepared realisations of the simpler homogeneous random walks, again with $\delta = \tau = 1$, we record the exit time and superimpose the histogram of the exit time distribution on the results for the true heterogeneous problem in Figure~\ref{fig:3} where we see that the histograms for the effective homogenised disc and sphere are very similiar to those of the true heterogeneous disc and sphere.  Additional results for the inward configuration or different arrangements of heterogeneous layers give rise to similar results (not shown).

\section{Conclusion}
In this work we consider random walk models of diffusion in heterogeneous environments with the aim of constructing, and validating, new exact formulae for exit time properties and homogenization.  Most exact results in the literature correspond to problems in relatively simple geometries with homogeneous material properties~\cite{Redner2001,Ellery2012a,Ellery2012b}.  Some consideration has been given to certain cases of heterogeneity~\cite{Carr2019,Kurella2014,Lindsay2015,Vaccario2015}.  For example Vaccario and colleagues~\cite{Vaccario2015} present exact expressions for the mean first passage time for a point particle diffusing in a spherically symmetric $d$-dimensional heterogeneous domains.  However, this previous study was limited to just two layers with one particular arrangement of boundary conditions. In contrast, our approach deals with an arbitrary number of layers, each layer of arbitrary thickness, and each layer with a distinct arbitrary value of the diffusivity.  Our approach is very flexible since we provide a framework that enables us to calculate exact expressions for any moment of the distribution of exit time and symbolic software is provided on GitHub\cite{Code} to facilitate such computations.  With such exact expressions we calculate a new suite of homogenization results that allow us to approximate some particular heterogeneous system with an effective homogeneous system with diffusivity $D_{\mathrm{eff}}$.

There are many possible extensions of the work presented here.  For example, the homogenization formulae in Table~\ref{Table4} are based on approximating a heterogeneous problem with an effective homogeneous media such that the first moment of exit time in the homogenised problem is identical to the first moment of exit time in the heterogeneous problems.  There are other ways that one could define an effective medium.  For example, another approach is to constrain $D_{\mathrm{eff}}$ in such a way that the homogenised system also accounts for higher moments of the exit time distribution~\cite{Carr2019}. Another extension would be to consider the exit time distributions for growing lines, discs and spheres that has been considered previously in the case of diffusion through homogeneous materials~\cite{Simpson2015,Simpson2015b}, but not through heterogeneous materials.

\section*{Supplementary Material}
See the supplementary material for additional computational results that confirm the accuracy of the moment expressions derived in section \ref{sec:moments}.

\bigskip
\noindent
\textit{Acknowledgements}. 
This work is supported by the Australian Research Council (DP200100177). We thank the anonymous reviewer for their helpful comments that improved the quality of the final manuscript.

%
\noindent
\textit{Data Availability}.
The data that support the findings of this study are openly available on GitHub at \href{https://github.com/elliotcarr/Carr2020c}{https://github.com/elliotcarr/Carr2020c}.

\end{document}